\documentclass[a4paper,11pt]{article}
\usepackage{jheppub} 
\usepackage[utf8]{inputenc}
\usepackage{physics}
\usepackage{slashed}
\usepackage{caption}
\usepackage{xcolor}
\usepackage{comment}
\usepackage{multirow}
\usepackage{graphics}
\usepackage{float}
\usepackage{cases}
\usepackage{cancel}
\usepackage{soul}
\usepackage{array}
\usepackage{mathtools}  
\usepackage{amsfonts}
\usepackage{hyperref}
\usepackage{amsmath}
\usepackage{amssymb}
\usepackage{tcolorbox}
\usepackage{tikz}     
\usepackage{newunicodechar}
\newunicodechar{−}{-}

\title{\boldmath\boldmath Exact Color–Kinematics Duality in the Cosmological Correlators}

\author{Aswini Bala, Sachin Jain, Vibhor Singh}

\affiliation{Indian Institute of Science Education and Research,\\ Dr Homi Bhabha Road, Pashan, Pune, India}

\emailAdd{aswini.bala@students.iiserpune.ac.in}
\emailAdd{sachin.jain@iiserpune.ac.in}
\emailAdd{singh.vibhorrajesh@students.iiserpune.ac.in}

\abstract{In this paper, we study color–kinematics duality and related double-copy and BCJ structures in the cosmological Grassmannian framework. The simplicity of these correlators in this framework makes it possible to test the double-copy construction beyond flat space. We show that the four-point gluon correlator admits a representation whose kinematic numerators satisfy an algebraically exact Jacobi identity, $n_s + n_t + n_u = 0$. Replacing color factors with kinematic numerators gives a graviton correlator that reproduces the correct factorization residues and flat-space limit, which matches with the BCFW result up to terms regular in all exchange channels. In addition, we construct a differential bi-adjoint scalar matrix whose null vector gives rise to a differential BCJ relation among color-ordered gluon correlators. We further explore an $s$-copy construction for general integer-spin correlators and show that for odd spin the corresponding numerators satisfy a Jacobi identity. Finally, we develop a differential representation for scalar four-point correlators with gluon exchange. These results establish algebraic and differential aspects of graviton and gluon correlators in curved spacetime.}

\begin{document}

\maketitle

\section{Introduction}
Color-kinematics (CK) duality and the associated double-copy construction
uncover a remarkably deep and interesting connection between gauge theory and gravity
\cite{Bern:2008qj,Bern:2022wqg}. CK duality/ double copy suggests that the
apparent complexity of gravity   may be understood using a simpler gauge-theory
structure. The double copy not only applies to tree-level but also extends well beyond to construct multiloop gravity integrands and it also helps to understand unexpected ultraviolet cancelations and the ultraviolet properties of supergravity theories \cite{Bern:2012uf, Bern:2012cd, Bern:2013uka, Bern:2017ucb, Bern:2018jmv}. It is interesting to note that, in classical gravity: black-hole and radiative solutions in suitable cases can be related to much simpler gauge-theory configurations
\cite{Monteiro:2014cda, Luna:2015paa, Luna:2016due, Luna:2016hge}. Other applications include studying observables relevant to binary dynamics and gravitational-wave physics
\cite{Buonanno:2022pgc, Kosower:2022yvp}. These developments point towards a deep structural property of gravity and its connection to gauge theory, rather than an accidental property of  some particular class of amplitudes. All these developments make it exciting to explore similar features in   curved spacetime, where it may provide a new organizing principle for gravitational observables in AdS and cosmology.

There has been some recent progress in understanding  color-kinematics duality and the double copy in (A)dS and CFT correlators. In momentum space, double-copy structures have been found for three-point CFT$_3$ correlators, while four-point A(dS)$_4$ correlators exhibit color-kinematics and double-copy relations with curvature-dependent modifications
\cite{Farrow:2018yni, Jain:2021qcl, Armstrong:2020woi,Albayrak:2020fyp,Armstrong:2023phb,Mei:2023jkb,Gomez:2026yno}. Independently, a differential
formulation in AdS has led to an analogous CK duality relation and  BCJ
relations in terms of conformal Casimir operators in the embedding space formulation \cite{Diwakar:2021juk,Herderschee:2022ntr}.

The cosmological Grassmannian \cite{Arundine:2026fbr,De:2026shn,Bala:2026hdm,Bala:2026bdx,Huang:2026tsh,Arundine:2026myr,Bala:2026lvw,Arundine:2026qqg} provides a particularly simple setting where both algebraic and differential relations can be investigated within a common framework.
The four-point correlators in this formalism reduce to remarkably
simple rational functions of Grassmannian Mandelstam variables and satisfy
factorization rules that are closely analogous to those of scattering amplitudes
\cite{Arundine:2026fbr}. More recently, a BCFW-like recursion was developed
directly in the cosmological Grassmannian, which provides  a purely algebraic
construction of the four-point correlators. For example, this construction yielded a simple result for the graviton four point function 
\cite{Bala:2026lvw}. This simplicity allows us to address the gauge-gravity relation algebraically. Interestingly,  it is easy to obtain Grassmannian expressions using BCFW recursion  for general spin-$s$ exchange, which naturally leads to the question of whether the usual double copy is the spin-two example  of a more general $s$-copy structure. Remarkably, the Grassmannian correlators also admit a differential description in terms of conformal Casimir operators. The Grassmannian therefore provides a unified framework in which algebraic CK duality, double- and $s$-copy relations, as well as differential BCJ relations, can all be studied together.

For example, for the four-gluon correlator consistent with the KK relation, we identify channel numerators, that satisfy the exact kinematic Jacobi identity, in spite of the fact that  Grassmannian Mandelstam variables obey $S+T+U\neq0$. The associated generalized gauge freedom, which can shift the numerators,  is essentially fixed once we impose the Jacobi condition. If we replace color factors with kinematic numerators, we obtain the Grassmannian double copy, which  correctly reproduces the factorization residues and flat-space limit of the graviton correlator. However, the correlator obtained by the double copy differs from the BCFW result by terms that are regular in the exchange channels. We then compute the spin-$s$ four point function with spin-$s$ exchange using BCFW recursion  and show, in particular, that odd-spin numerators again satisfy a Jacobi identity. The corresponding $s$-copy correlator obtained using the s-copy of the YM kinematic numerator has also the correct  factorization property. 

 We also construct a differential bi adjoint-scalar matrix whose null vector yields a differential BCJ relation. 
These results demonstrate that the cosmological Grassmannian preserves much of the algebraic structure of flat-space color-kinematics duality, at the same time, it  displays features of curved space results in a simpler setting.

This paper is structured as follows: In \textit{section \ref{section2}}, we review the basic ingredients of color-kinematics duality, BCJ relations, and the double copy in flat space. We then  summarize their extensions to AdS correlators. In \textit{section \ref{section3}}, using cosmological Grassmannian framework, we show four gluon correlator numerator satisfies  Jacobi identity. We also analyze the corresponding graviton double copy, and extend the construction to general spin through the $s$-copy. In \textit{section \ref{section4}}, we discuss the Grassmannian biadjoint-scalar matrix and develop its differential counterpart. We then use this matrix to show differential BCJ relation. 
In \textit{section \ref{section5}}, we present a differential representation for the four-scalar
correlator with gluon exchange. We conclude in \textit{section \ref{section6}} with a brief discussion of the
results and future directions.

\section{Double Copy and CK duality : A Review}\label{section2}

There have been many developments in understanding the color-kinematics (CK) duality, BCJ relations, KLT relations, and the double-copy construction in both flat and anti-de Sitter (AdS) space. In this section, we provide a brief review of these developments, pointing to the structures relevant to our discussion. For an extensive review of the subject in flat space and AdS refer \cite{Bern:2019prr, CarrilloGonzalez:2026phk}.

\subsection{Flat Space Double Copy}
\label{sec:flat-review}

The Color-Kinematic duality was first proposed for flat space amplitudes in \cite{Bern:2008qj}. At tree level, the four-point Yang-Mills amplitude can be written in terms of color and kinematic structures as
\begin{equation}
    \mathcal{A}_4^{gluon} = \frac{c_s n_s}{s} + \frac{c_t n_t}{t} + \frac{c_u n_u}{u}
    \label{eq:YM-flat}
\end{equation}
where $c_i$ are the color factors made out of the structure constants of the gauge group, satisfying the Jacobi identity,
\begin{equation}
    c_s+c_t+c_u = 0
\end{equation}
and $n_i$ are the kinematic numerators. For four-point scattering of massless particles in flat space, momentum conservation and the on-shell conditions imply,
\begin{equation}
    s+t+u=0.
    \label{eq:flat-mandelstam}
\end{equation}
This kinematic constraint allows the numerators to be chosen such that they obey the Jacobi identity,
\begin{equation}
    n_s + n_t + n_u = 0.
    \label{eq:flat-Jacobi}
\end{equation}
Thus allowing the color and kinematic factors to satisfy the same algebraic properties, establishing a color-kinematic duality. This duality allows us to relate the Yang-Mills amplitudes to the gravity amplitudes through double copy \cite{Bern:2008qj} by replacing the color factors in \eqref{eq:YM-flat} to the kinematic numerators $c_i \to n_i$, giving 
\begin{equation}
    \mathcal{A}^{grav}_4 = \frac{n_s^2}{s} + \frac{n_t^2}{t} + \frac{n_u^2}{u}.
\end{equation}
Note that in the Yang-Mills amplitude \eqref{eq:YM-flat}, if we perform a generalized gauge transform $(n_s,n_t,n_u)\to (n_s',n_t',n_u')$,
\begin{equation}\label{eq:generalizedgauge}
    n_s' = n_s + s\chi, \quad n_t'= n_t + t \chi, \quad n_u' = u \chi
\end{equation}
full colored-amplitude and the Jacobi identity remain invariant, i.e.
\begin{align}
    &\mathcal{A}_4^{gluon} = \frac{c_s n_s}{s} + \frac{c_t n_t}{t} + \frac{c_u n_u}{u} \to \frac{c_s n_s}{s} + \frac{c_t n_t}{t} + \frac{c_u n_u}{u} + \chi (c_s +c_t +c_u) = \mathcal{A}_4^{gluon},\notag\\
    & n_s + n_t +n_u\to n_s + n_t + n_u +\chi (s+t+u) = n_s + n_t + n_u.
\end{align}
\textbf{Bi-adjoint Scalar and BCJ relation}

\noindent We can get a zeroth copy by replacing the kinematic factors in \eqref{eq:YM-flat} to color factors $n_i\to \tilde{c}_i$, giving the amplitude for a bi-adjoint scalar theory which transforms under two adjoint representations \cite{Cachazo:2013iea,CarrilloGonzalez:2026phk},
\begin{equation}
    \mathcal{M}^{(0)}_4 = \frac{c_s \tilde{c}_s}{s} + \frac{c_t \tilde{c}_t}{t} + \frac{c_u \tilde{c}_u}{u}.
\end{equation}
We can write the double color ordered amplitudes for the biadjoint scalar in the DDM basis \cite{CarrilloGonzalez:2026phk} $(1234,1324)$ in matrix form (BAS Matrix) as
\begin{equation}
    m_4 = \begin{pmatrix} \frac{1}{s} + \frac{1}{t} & -\frac{1}{t}\\ -\frac{1}{t} & \frac{1}{t} + \frac{1}{u} \end{pmatrix}.
    \label{eq:flat-BAS-matrix}
\end{equation}
Here, we can notice that since the flat space kinematics gives $s+t+u=0$, this matrix has rank-1, and thus allows for a null vector $v = (s,-u)$, which can be used to relate the two color-ordered Yang-Mills amplitudes, giving the BCJ relation \cite{Vaman:2010ez},
\begin{equation}
    s A_{ST} = u A_{TU},
\end{equation}
where $A_{ST}$ and $A_{TU}$ are the color-ordered Yang-Mills amplitudes given by,
\begin{equation}
    A_{ST} = \frac{n_s}{s}-\frac{n_t}{t}, \quad A_{TU} = \frac{n_t}{t} - \frac{n_u}{u}.
    \label{eq:flat-color-ordered}
\end{equation}
It is interesting to note that the biadjoint-scalar (BAS) matrix also provides a natural way to the KLT
relations, which relates Yang Mills and gravity amplitudes \cite{CarrilloGonzalez:2026phk}. For example, at tree level four points, the KLT relation \cite{Elvang:2013cua} is,
\begin{equation}
    {\cal A}_4^{grav}= -s A_{4,ST}^{gluon} A_{4,US}^{gluon}. 
\end{equation}
The KLT kernel s is obtained as a restricted inverse of BAS matrix on its non-null subspace \cite{Mizera:2016jhj}.
Thus, the BAS matrix provides an efficient and unified way of relating biadjoint-scalar,
Yang--Mills and gravitational amplitudes. 

\subsection{AdS Space Double Copy}
\label{sec:AdS-review}
Due to the absence of an s-matrix description in AdS space, the natural observables are the boundary CFT$_3$ correlation functions. They are described in terms of the boundary momenta $\mathbf{k}_i$ or equivalently in 3d spinor-helicity variables as $k_{i,a b} = \lambda_{(a}\bar{\lambda}_{b)}$. Due to different kinematic structures, the CK duality construction does not generalize to AdS trivially. Nevertheless, there have been several developments in understanding color-kinematics duality and double copy structures for $AdS_4$ correlators \cite{Armstrong:2020woi, Diwakar:2021juk,Herderschee:2022ntr,Cheung:2022pdk, CarrilloGonzalez:2026phk}. We give a brief review of them below.\footnote{The double copy for AdS$_6$ has been studied in \cite{Albayrak:2020fyp}}, with the first part of the review following very closely \cite{Armstrong:2020woi}.

\noindent \textbf{Momentum Space :} Let us define the Mandelstam variables,
\begin{align}
    k_s = (k_1  + k_2 + |\textbf{k}_1 + \textbf{k}_2|)(k_3  + k_4 + |\textbf{k}_3 + \textbf{k}_4|)\notag\\
    k_t = (k_1  + k_4 + |\textbf{k}_1 + \textbf{k}_4|)(k_2  + k_3 + |\textbf{k}_2 + \textbf{k}_3|)\notag\\
    k_u = (k_1  + k_3 + |\textbf{k}_1 + \textbf{k}_3|)(k_2  + k_4 + |\textbf{k}_2 + \textbf{k}_4|)
\end{align}
where $s+t+u\neq 0.$
The color-ordered correlators can be written analog to flat space as ,
\begin{equation}
    \mathcal{A}(1234) = \frac{n_s}{k_s} - \frac{n_t}{k_t}, \quad \mathcal{A}(1324) = \frac{n_t}{k_t} - \frac{n_u}{k_u}.
\end{equation}
Note that the kinematic numerators defined here do not satisfy the Jacobi identity,
\begin{equation}
    n_s + n_t +n_u \neq 0.
\end{equation}
One can use  gauge transformation, $(n_s,n_t,n_u)\to (n_s',n_t',n_u')$,
\begin{equation}
    n_s'= n_s + k_s \chi, \quad n_t'= n_t + k_t \chi, \quad n_u'= n_u + k_u \chi
\end{equation}
and demanding the Jacobi identity $n_s'+n_t'+n_u'=0$ in new variables would give $\chi = -\frac{n_s+n_t+n_u}{k_s+k_t+k_u}.$ However, one can check that, since $k_s+k_t+k_u\neq 0$, the BCJ relation is not satisfied and is given by 
\begin{equation}
    u \mathcal{A}(1324) - s \mathcal{A}(1234) = (s+t+u)\frac{n_t'}{t} = \xi \frac{n_t'}{t}
\end{equation}
where $\xi = s+t+u$.

For the four-graviton correlator in AdS$_4$, the ususal  flat-space double-copy prescription does not work, it requires an important modification.  Once we re arrange the Yang--Mills kinematic numerators to satisfy the Jacobi relation, the naive CK square reproduces the correct flat-space limit. However, it does not by itself give the full AdS$_4$ correlator and contains spurious singularities absent in the physical answer.  The physical correlator therefore takes the schematic form
\begin{equation}
\langle TTTT\rangle_{\rm AdS} =
\langle TTTT\rangle_{\rm DC} +
\Delta_{\rm AdS},
\end{equation}
where $\langle TTTT\rangle_{\rm DC}$ denotes the contribution obtained by squaring the gauge-theory kinematic data. The $\Delta_{\rm AdS}$ is a curvature-dependent local/contact completion which is required to cancel spurious singularities and restore the full AdS consistency conditions.  It is worth emphasizing that  $\Delta_{\rm AdS}$ does not modify the flat-space amplitude extracted from the total-energy singularity, so the ordinary gravitational double copy is recovered in the flat-space limit \cite{Armstrong:2023phb,CarrilloGonzalez:2026phk}.

\paragraph{Differential CK duality and double copy in AdS for four external scalar with gluon and graviton exchange.}
A useful formulation of color--kinematics duality for AdS boundary correlators is obtained in position space, in embedding coordinates  by replacing flat-space Mandelstam invariants by quadratic conformal-Casimir operators,
$\mathbb D_{ij}\equiv(D_i+D_j)^2$, acting on a common contact Witten diagram $D_{\Delta_1\Delta_2\Delta_3\Delta_4}$. The inverse Casimir then play  role of propagators \cite{Diwakar:2021juk, Herderschee:2022ntr}.  If we consider the case of four adjoint scalars with  gluon exchange, we get
\begin{equation}
{\cal A}_4=
\left(
C_s\,\frac{\widehat N_s}{\mathbb D_{12}}
+C_t\,\frac{\widehat N_t}{\mathbb D_{23}}
+C_u\,\frac{\widehat N_u}{\mathbb D_{13}}
\right)D_{d,d,d,d},
\qquad
\widehat N_s=\mathbb D_{23}-\mathbb D_{13},
\end{equation}
with cyclic permutations for $\widehat N_{t,u}$ and
$\widehat N_s+\widehat N_t+\widehat N_u=0$.  The factors $C_s$ are the usual color factors. 

\noindent Similarly, we can define a differential analogue of the BAS matrix as
\begin{equation}
    \hat{m}_4 = \begin{pmatrix} \mathbb{D}_{12}^{-1} + \mathbb{D}_{23}^{-1} & -\mathbb{D}_{23}^{-1} \\
-\mathbb{D}_{23}^{-1} & \mathbb{D}_{13}^{-1} + \mathbb{D}_{23}^{-1} \end{pmatrix},
\end{equation}
which admits a null vector $\left(\mathbb{D}_{12},-\mathbb{D}_{13}\right)\hat{m}_4 \simeq 0$ giving the Differential BCJ relation for Scalar with gluon exchange correlator
\begin{equation}
    \mathbb{D}_{12}\mathcal{A}(1234)=\mathbb{D}_{13}\mathcal{A}(1324)
\end{equation}
providing a direct operator analogue of $sA(1234)=uA(1324)$ in flat space \cite{Diwakar:2021juk,Herderschee:2022ntr}.  The corresponding scalar correlator with graviton exchange shows both the usefulness as well as   the limitation of the naive double copy:
\begin{equation}\label{diffdub}
{\cal M}_4=
\frac1{16}\sum_{i=s,t,u}
\frac{\widehat N_i^{\,2}+2d\,\mathbb D_i}
     {\mathbb D_i+2d}\,
D_{d,d,d,d},
\end{equation}
so that in AdS$_4$ ($d=3$) the replacement is schematically
$\widehat N_i^2/\mathbb D_i\rightarrow
(\widehat N_i^2+6\mathbb D_i)/(\mathbb D_i+6)$.
We conclude that the gauge-theory numerator square determines the leading factorizing structure. The finite AdS curvature generates additional Casimir-dependent terms which appear in \eqref{diffdub}. The ordinary flat-space double copy is recovered when these curvature corrections are removed in the flat-space limit \cite{Herderschee:2022ntr}.
\paragraph{Differential BCJ relation}
\noindent In \cite{Diwakar:2021juk,Herderschee:2022ntr} it was also shown that for the  Yang--Mills four-current  color-ordered correlators obey the analogous differential BCJ relation
\begin{equation}
\mathbb D_{12}{\cal A}_{\rm YM}(1234)
=
\mathbb D_{13}{\cal A}_{\rm YM}(1324).
\end{equation}
 The explicit four-current correlator check there was performed for a four-dimensional boundary.  However, at four points gauge-to-gravity differential double copy for the four-stress-tensor correlator is not very well understood \cite{Diwakar:2021juk, Herderschee:2022ntr,CarrilloGonzalez:2026phk}. 
 
 In the following sections, we revisit these issues within the cosmological Grassmannian framework and show how color–kinematics duality, BCJ relations, and the double-copy construction are realized in this setting.

 \section{Double Copy and CK duality in Grassmannian}\label{section3}
Having discussed the Flat space and AdS space developments in CK duality and double copy, we want to explore these aspects in the Cosmological Grassmannian space. The simplicity of correlation functions in the Grassmannian space allows us to test these aspects in efficient manner. We begin by reviewing the Grassmannian for the 3d CFT correlators. 

The idea is to embed the kinematical data ($\Lambda$) of the external operators into a null plane (C) called the orthogonal Grassmannian in the kinematical space. Then the conformal correlators are represented by \cite{Arundine:2026fbr}:
\begin{align} \label{eq:grassrep}
\Psi_n^{h_1h_2\cdots}(\Lambda)=\int {d^{n\times 2n}C\over Vol(\mathbb{GL}(n))}\,\delta(C\cdot Q\cdot C^T)\delta(C\cdot \Lambda)\, A_n^{h_1h_2\cdots}(C),
\end{align}
where Q is the metric of $\mathbb R^{n,n}$, whereas $A_n(C)$ contains all the dynamical information and it is represented as a function of $(n\times n)$ minors of the C matrix. The C matrix is labeled by their columns, i.e. $C = \{\bar1\bar2\cdots \bar n 12\cdots n\}$. The Grassmannian integral solves the homogeneous Ward identity and produces a particular discontinuity, and the full correlator can be reconstructed by a dispersive integral. For our case, we will be focusing on four point correlators, for which we introduce some notation that will be used henceforth.  

 In the orthogonal Grassmannian description, there are two disconnected spaces, right and left branch. We choose the right-branch gauge of OGr(4,8) to represent our correlators,
\begin{equation}
    C = (\mathbb{I}_{4\times 4},-c_{4\times 4}), \qquad c_{ij} = -c_{ji}
\end{equation}
and define the Mandelstam variables in the Grassmannian space as minors,
\begin{align}
    S = (\bar{1}\bar{2}12) = c_{13}c_{24} - c_{14}c_{23},~ T = (\bar{1}\bar{4}14) = c_{13}c_{24} - c_{12}c_{34},~ U =(\bar{1}\bar{3}13) = -c_{14}c_{23} - c_{12}c_{34}.
\end{align}
Note that, unlike flat-space Mandelstam’s, they do not satisfy $S+T+U = 0$. We also define the following quantities,
\begin{equation}
    \Sigma = S+T+U, \quad \Delta_S = -S+T+U, \quad \Delta_T = S-T+U, \quad \Delta_U = S+T-U.
\end{equation}

\subsection{Gluon CK Duality}
In Grassmannian space, using the BCFW bridge \cite{Bala:2026lvw}, we can construct the \textit{reduced} and \textit{complete} color-ordered gluon correlator \cite{Arundine:2026fbr} for Yang-Mills theory, which when converted to momentum space corresponds to double and triple discontinuities of the Euclidean correlators, respectively.

 The reduced four-point color ordered gluon correlators with a gluon exchange for the case of our interest are given by,
\begin{align}
    \hat{A}_{ST}^{--++} &=\hat A(1234)= 2\frac{H}{ST}\left( \frac{1}{\Sigma} + \frac{1}{\Delta_U}\right), \notag \\  
    \hat{A}_{TU}^{-++-}&=\hat A(1342) = 2\frac{H}{TU}\left( \frac{1}{\Sigma} + \frac{1}{\Delta_S}\right), \notag \\
    \hat{A}_{US}^{-+-+} &= A(1423)= 2\frac{H}{US}\left( \frac{1}{\Sigma} + \frac{1}{\Delta_T}\right),
\end{align}
where $H = (12\bar{3}\bar{4})^2$.
Note that if we sum up the above reduced color-ordered correlators, the Kleiss–Kuijf (KK) relation is not satisfied \cite{Arundine:2026fbr},
\begin{equation}
\hat{A}_{ST}+\hat{A}_{TU}+\hat{A}_{US} = \frac{8H}{\Delta_S \Delta_T \Delta_U}\neq 0.
\end{equation}
\noindent However, if we use the complete color-ordered correlators 
\begin{align}
    A_{ST}^{--++} &= \frac{H}{ST}
    \left[
        \frac{3}{\Sigma}
        + \frac{1}{\Delta_U}
        - \frac{1}{\Delta_S}
        - \frac{1}{\Delta_T}
    \right], \notag \\
    A_{TU}^{-++-} &= \frac{H}{TU}
    \left[
        \frac{3}{\Sigma}
        + \frac{1}{\Delta_S}
        - \frac{1}{\Delta_T}
        - \frac{1}{\Delta_U}
    \right],\notag \\
    A_{US}^{-+-+} &= \frac{H}{US}
    \left[
        \frac{3}{\Sigma}
        + \frac{1}{\Delta_T}
        - \frac{1}{\Delta_U}
        - \frac{1}{\Delta_S}
    \right].
\end{align}
they do satisfy the KK relation \cite{Arundine:2026fbr}, $A_{ST}+A_{TU}+A_{US}=0$.

Our particular interest is the gluon correlator, which satisfies the KK relation for the reasons that will be discussed below.
Rewriting the complete colored ordered correlators as,
\begin{equation}\label{eq:coloredYM}
    A_{ST}^{--++} =  \frac{n_s}{S} - \frac{n_t}{T}, \quad A_{TU}^{-++-} = \frac{n_t}{T} - \frac{n_u}{U}, \quad A_{US}^{-+-+} = \frac{n_u}{U} - \frac{n_s}{S},
\end{equation}
we can read-off the kinematic numerators
\begin{equation}
    n_s = -\frac{4H(T-U)}{\Sigma \Delta_T\Delta_U}, \quad n_t = -\frac{4H(U-S)}{\Sigma \Delta_S\Delta_U}, \quad n_u = -\frac{4H(S-T)}{\Sigma \Delta_T\Delta_S}.
    \label{eq:Grass-numerators}
\end{equation}
Using the above equations, the full gluon color-dressed correlator can be written as,
\begin{align}\label{eq:fullYMcorrelator}
    A_4^{gluon} = \frac{n_s c_s}{S}+\frac{n_t c_t}{T}+\frac{n_u c_u}{U}.
\end{align}

A remarkable feature of the above obtained kinematic factors is that they satisfy the kinematic Jacobi identity
\begin{equation}
    n_s+n_t+n_u=0,
    \label{eq:KinematicJacobi}
\end{equation}
and hence establishes an algebraic kinematics Jacobi identity \footnote{{Please note that a differential color kinematics duality was obtained in mellin-momentum space \cite{Mei:2023jkb}, where as we are establishing here an algebraic identity.} }. 

Also note that given the expressions in \eqref{eq:coloredYM} and \eqref{eq:Grass-numerators}, the naive generalization of the algebraic flat space BCJ relation does not hold eq \eqref{determinantBAS}.
This will be discussed in detail in subsequent sections.

\paragraph{Generalized gauge freedom in the cosmological Grassmannian.}
  Let us note that, there is a direct analogue of the usual generalized gauge transformation eq \eqref{eq:generalizedgauge}, in cosmological Grassmannian,
\begin{equation}
    n_s\rightarrow n_s+S\,\chi,\qquad
    n_t\rightarrow n_t+T\,\chi,\qquad
    n_u\rightarrow n_u+U\,\chi ,
\end{equation}
where $\chi=\chi(S,T,U)$ is arbitrary.  This leaves the correlator \eqref{eq:fullYMcorrelator} unchanged as
\begin{equation}
    \delta{\cal A}_4
    =\chi\,(c_s+c_t+c_u)=0.
\end{equation}
On the other hand, the kinematic Jacobi combination transforms as
\begin{equation}
    n_s+n_t+n_u
    \longrightarrow
    n_s+n_t+n_u+\Sigma\,\chi .
\end{equation}
Hence, for $\Sigma\neq0$, one may choose
\begin{equation}
    \chi=-\frac{n_s+n_t+n_u}{\Sigma}
\end{equation}
to reach a color--kinematics representation satisfying
\begin{equation}
    n_s+n_t+n_u=0.
\end{equation}
An important consequence is that, once we impose  Jacobi condition , the residual transformation must obey $\Sigma\chi=0$ and is therefore trivial for generic cosmological kinematics as $\Sigma\neq 0$.

\subsection{Graviton double copy}

Now, from the Color-Kinematic duality, if we take the Yang Mills correlator \eqref{eq:fullYMcorrelator} and replace the color factors with kinematic numerators, we expect to obtain the four-point graviton correlator,
\begin{align}
    A_4^{grav} = \frac{n_s^2}{S} + \frac{n_t^2}{T} + \frac{n_u^2}{U}= \frac{4H^2}{\Sigma^2}
    \left[
        \frac{(q_U-q_T)^2}{S}
        +
        \frac{(q_S-q_U)^2}{T}
        +
        \frac{(q_T-q_S)^2}{U}
    \right],
\end{align}
where,
\begin{equation}
    q_S = \frac{1}{\Delta_S}, \quad q_T = \frac{1}{\Delta_T}, \quad q_U = \frac{1}{\Delta_U}.
    \label{eq:q-variables}
\end{equation}
Whereas the Grassmannian graviton 4 point correlator obtained using the BCFW bridge \cite{Bala:2026lvw} is given by,
\begin{equation}\label{eq:gravBCFW}
    {A}_{4}^{grav, BCFW} = \frac{8H^2}{\Sigma^2}
    \left[
        \frac{q_U^2 + q_T^2}{S}
        +
        \frac{q_U^2 + q_S^2}{T}
        +
        \frac{q_T^2 + q_S^2}{U}
    \right].
\end{equation}
Hence, the difference $A^{\mathrm{diff}}$ is,
\begin{equation}\label{Adiff1}
    A^{\mathrm{diff}} = A_4^{g}-{A}_{4}^{grav, BCFW} = \frac{16H^2}{\Sigma^2}
    \left[\frac{S}{\Delta_T^2 \Delta_U^2}+\frac{T}{\Delta_S^2 \Delta_U^2}+\frac{U}{\Delta_S^2 \Delta_T^2}\right],
\end{equation}
the difference is regular in $S,T, U,$ and thus, ${A}_{4}^{grav, BCFW}$ and $A_4^{g}$ agree with each other up to a regular (contact) term. Hence, they both match at the factorization of the exchange channel. 

The flat space limit for ${A}_{4}^{grav, BCFW}$ and $A_4^{g}$ also matches,
\begin{equation}
\lim_{E\to 0}A_4^{grav}=\lim_{E\to 0}A_4^{grav, BCFW}
\sim -\frac{1}{E^3}\frac{\langle12 \rangle^4 [34]^4}{stu}.
\end{equation}
It is also interesting to note that, even if the $A^{\text{diff}}$ in \eqref{Adiff1} is regular in the exchange minors S, T and U, but has the same flat space limit as to the graviton 4 point. 
So it would be interesting, to compare  the $A_4^{grav}$, obtained from double copy  with the momentum space known graviton answer. 
\subsection{Comments on the $s$-copy}
Here, we give some comments about the $s$-copy generalization of the $\langle JJJJ\rangle$ double copy. We consider the all equal external integer spin-$s$ four point correlator $\langle J^-_s(1)J^-_s(2)J^+_s(3)J^+_s(4)\rangle$ with $J_s$ exchange.

Taking the three-point leading interaction of the Cosmological Grassmannian \cite{Arundine:2026fbr} as the input and using the multiple BCFW bridge deformation \cite{Bala:2026lvw} to construct the four-point spin-$s$ correlator in $(--++)$ helicity, we obtain
\begin{equation}
    {A}_{4}^{J_s, BCFW} = \frac{N_S^{(s)}}{S} + \frac{N_T^{(s)}}{T}+\frac{N_U^{(s)}}{U} + B_{\infty}^{(s)},
\end{equation}
where the spin-$s$ kinematic numerators are given by
\begin{align}
    N_S^{(s)}=\frac{1}{2}\left(\frac{4H}{\Sigma}\right)^s\left( q_U^s +(-1)^s q_T^s\right), N_T^{(s)}=\frac{1}{2}\left(\frac{4H}{\Sigma}\right)^s\left( q_S^s +(-1)^s q_U^s\right),N_U^{(s)}=\frac{1}{2}\left(\frac{4H}{\Sigma}\right)^s\left( q_T^s +(-1)^s q_S^s\right),
\end{align}
with $q_S, q_T$ and $q_U$ defined in \eqref{eq:q-variables}, $H=(12\bar3\bar4)^2$ and $B_{\infty}^s$ denote the boundary term invisible to the deformation. 

One interesting thing to note that, for odd integer $s$, these numerators satisfy,
\begin{equation}
    N_S^{(s)} + N_T^{(s)} + N_U^{(s)} = 0.
\end{equation}
So, they admit a generalization of the adjoint CK duality type relation, and for even $s$, the construction is gravity-like. 

We compare the correlators obtained from the BCFW relation to the $s$-copy of the spin-one CK representation, which is given by,
\begin{equation}
    {A}_{4,J_s}^{s-copy}\equiv \frac{n_S^s}{S} + \frac{n_T^s}{T} + \frac{n_U^s}{U}.
\end{equation}
 $n_S,n_T$ and $n_U$ are the kinematic numerators of the Yang-Mills correlator given in \eqref{eq:Grass-numerators}. 

Now, taking the difference of the kinematic numerator for the S-channel of $A_{4,J_s}^{s-copy}$ and $A_{4,J_s}^{BCFW}$, 
   \begin{align}\label{eq:diffscopy}
       A_{S,diff}^{(s)}=N_S^{(s)} - n_S^{(s)}= 4 \bigg(\frac{2 H}{\Sigma}\bigg)^s \frac{S}{\Delta_U^2 \Delta_T^2} Q_{s-2}\bigg(\frac{1}{\Delta_U},\frac{1}{\Delta_T}\bigg), 
   \end{align}
where,
 \begin{align}
     Q_{s-2}(x,y)= \sum_{r=1}^{\left\lfloor s/2 \right\rfloor} \binom{s}{2r} (x-y)^{s-2r}(x+y)^{2r-2}, ~\forall~ s\geq 2.
 \end{align}
 One can note that the difference \eqref{eq:diffscopy} is regular in the exchange channel S. Hence, the s-copy correlator will match at the factorization.\footnote{The difference polynomial $Q_{s-2}$ that we have got has an interesting interpretation suggested by OpenAI. Writing the spin-one numerator, say the s-channel, \(n=u+v\), where $u=\frac{1}{\Delta U}, v=\frac{-1}{\Delta T}$. If \(u\) and \(v\) are assigned two opposite spin-one weights, i.e. u:h=+1,v:h=−1. The naive power \(n^s\) contains, besides the extremal helicities \(\pm s\), lower-helicity sectors \(s-2,s-4,\ldots\), whereas
the physical massless spin-\(s\) projector  keeps only \(u^s+v^s\) as given by BCFW. With \(m=x-y\) and \(d=x+y\), this projection gives
\[
m^s+d^2Q_{s-2}(m,d)
=\frac12\big[(m+d)^s+(m-d)^s\big]
=(m^2-d^2)^{s/2}
T_s\!\left(\frac{m}{\sqrt{m^2-d^2}}\right),
\] where $T_s$ is the Chebyshev polynomial of the first kind.
Hence, \[
Q_{s-2}(m,d)=
\frac{(m^2-d^2)^{s/2}
T_s\!\left(m/\sqrt{m^2-d^2}\right)-m^s}{d^2}.\] 
Thus, $Q_{s-2}$ is precisely the lower-degree harmonic (trace) tail that converts the naive s-fold product into the $\pm s$ projection.} 
\section{Differential BCJ relation in Grassmannian}\label{section4}
In this section, we show that the naive generalization of the flat space BCJ relation does not carry over to Grassmannian space, establish a differential representation for the BAS matrix and derive a differential BCJ relation. We also give a Grassmannian extension of the differential construction of scalar correlators with gluon exchange \cite{Diwakar:2021juk,Herderschee:2022ntr} (refer to section \ref{sec:AdS-review} for a brief review).
\subsection{Bi-Adjoint Scalar}
In flat space, the BAS matrix (refer sec (\ref{sec:flat-review})) of the bi-adjoint scalar theory gives us the null vector required to establish the BCJ relation.

To establish this connection in Grassmannian, we first introduce the BAS correlator in Grassmannian. 
The biadjoint scalar theory consists of scalar fields that transform under two adjoint representations of possibly different groups. It's Lagrangian in the bulk is given by \cite{CarrilloGonzalez:2026phk},
\begin{equation}
    \mathcal{L} = -\frac{1}{2}(\partial_{\mu}\phi^{a\tilde{a}})(\partial^{\mu}\phi^{a\tilde{a}}) - \frac{1}{3!}f^{abc}f^{\tilde{a}\tilde{b}\tilde{c}}\phi^{a\tilde{a}}\phi^{b\tilde{b}}\phi^{c\tilde{c}}.
\end{equation}
The four point boundary correlator for this theory can be computed using Witten diagrams. But one can obtain all the channels by solving the Casimir equation \cite{Arundine:2026myr}. In general, the solution will have homogeneous and non-homogeneous components. For external scalar with scalar exchange (S channel) for $\Delta = 2$ is given by \cite{Arundine:2026myr},
\begin{equation}
    A_{\Delta=2} = \frac{\textrm{Li}_2(1-2\frac{S}{\Sigma})-\pi^2/6}{2S},
\end{equation}
but for our purposes, we will only be interested in the homogeneous solution $1/S$. The fact that this is homogeneous can be seen from the action of the Casimir\footnote{The Casimir operator for the S,T and U channel for any conserved spin-s exchange, in Grassmannian is denoted by $\mathcal{C}_S^{[s]}, \mathcal{C}_T^{[s]}~ \text{and} ~ \mathcal{C}_U^{[s]} $ respectively.} \eqref{eq:Casimir12} for $l_1=l_2=0$. For the S-channel,
\begin{equation}
    (-\mathcal{C}_{s}^{[0]}+2)\frac{1}{S} = 0
\end{equation}
where the superscript $[0]$ denotes scalar external legs. A similar expression holds for T and U channels.

Thus, the full color-dressed homogeneous correlator of the BAS theory in the Grassmannian space is given by
\begin{equation}
\mathcal{A}_{4,\mathrm{BAS}}^{\mathrm{hom}} = \frac{c_s \tilde{c}_s}{S}
        +
        \frac{c_t \tilde{c}_t}{T}
        +
        \frac{c_u \tilde{c}_u}{U},
    \label{eq:BAScorrelator}
\end{equation}
where $c_i$ and $\tilde{c}_i$ are the color factors corresponding to the gauge group of the bi-adjoint scalar satisfying
\begin{equation}
    c_s + c_t + c_u = 0, \quad \tilde{c}_s + \tilde{c}_t + \tilde{c}_u = 0.
\end{equation} Note that the above BAS correlator \eqref{eq:BAScorrelator} can be seen as a zeroth copy of the Yang-Mills correlator \eqref{eq:fullYMcorrelator} where we replace the kinematic factor $\tilde{n}_i$ by $c_i$.

\noindent Given the full homogeneous BAS correlator eq \eqref{eq:BAScorrelator}, the double color-ordered correlator can be arranged in a BAS matrix,
\begin{equation}
    (m_4)_{\alpha\beta} = m_4[\alpha|\beta],
\end{equation}
where $\alpha$ and $\beta$ denote the color-ordering of the correlator. For $\alpha,\beta \in \{ 1234,1324 \}$, the BAS matrix is written as
\begin{equation}
    m_4^{\mathrm{hom}} = \begin{pmatrix} m_4[1234|1234] & m_4[1234|1324]\\ m_4[1324|1234] & m_4[1324|1324] \end{pmatrix} = \begin{pmatrix} \frac{1}{S}+\frac{1}{T} & -\frac{1}{T}\\ -\frac{1}{T} & \frac{1}{T}+\frac{1}{U} \end{pmatrix}.
    \label{eq:BAS-Grass}
\end{equation}
However, the algebraic flat space analog of the BCJ relation does not carry over to the Grassmannian space because the determinant of the BAS matrix \eqref{eq:BAS-Grass},
\begin{equation}\label{determinantBAS}
    \mathrm{det}(m_4^{\mathrm{hom}}) = \frac{S+T+U}{STU}\neq 0.
\end{equation}
Thus, this matrix has full rank and  does not have a null vector to establish the BCJ relation.


\subsection{Differential BCJ relation}\label{sec:DiffBCJ}
Above, we showed that the algebraic BAS matrix in the cosmological Grassmannian has no null vector for generic kinematics, since $(S+T+U\neq 0)$. Thus, the conventional flat-space BCJ relation does not hold, despite the exact kinematic Jacobi identity established in section \ref{sec:AdS-review}. Motivated by the differential formulation of AdS correlators \cite{Diwakar:2021juk}, we now construct a differential BAS matrix using conformal Casimir operators and investigate the resulting differential BCJ relation. Consider a differential representation of the BAS matrix \eqref{eq:BAS-Grass}, where we replace $S\to \mathcal{C}_s^{[1]}$, $T\to \mathcal{C}_t^{[1]}$ and $U\to \mathcal{C}_u^{[1]}$, and write the differential BAS Matrix as
\begin{equation}
    \hat{m}_4^{\mathrm{diff}} = \begin{pmatrix}
        \left(\mathcal{C}_s^{[1]}\right)^{-1} + \left(\mathcal{C}_t^{[1]}\right)^{-1}
        &
        -\left(\mathcal{C}_t^{[1]}\right)^{-1} \\[6pt]
        -\left(\mathcal{C}_t^{[1]}\right)^{-1} &
        \left(\mathcal{C}_u^{[1]}\right)^{-1} + \left(\mathcal{C}_t^{[1]}\right)^{-1}
    \end{pmatrix}.
\end{equation}
Since the Casimir\footnote{Let us note that $\mathcal{C}^1$ is Casimir, which acts on the internal spin-1 gluon correlator.}$\mathcal{C}_i^{[1]}$ eq \eqref{eq:Casimir12} satisfies
\begin{equation}
    \mathcal{C}_s^{[1]} + \mathcal{C}_t^{[1]} + \mathcal{C}_u^{[1]} = 0,
    \label{eq:diff-Jacobi}
\end{equation}
which is the analog of $s+t+u = 0$ in flat space. We observe  that the above differential BAS matrix has zero determinant as an operator and hence admits a null vector $v^T = (\mathcal{C}_s^{[1]}, -\mathcal{C}_u^{[1]})$,
\begin{align}
    \left(\mathcal{C}_s^{[1]},-\mathcal{C}_u^{[1]}\right)\hat{m}_4^{\mathrm{diff}}
    &=
    \left(
        1+(\mathcal{C}_s^{[1]}+\mathcal{C}_u^{[1]})(\mathcal{C}^{[1]}_t)^{-1},
        -1-(\mathcal{C}_s^{[1]}+\mathcal{C}_u^{[1]})(\mathcal{C}^{[1]}_t)^{-1}
    \right)\notag\\
    &= (0,0).
\end{align}
 We have used eq \eqref{eq:diff-Jacobi} to get the second line. Now, this null vector $v$ can be used to establish a BCJ relation among the gluon color-ordered correlators as 
\begin{equation}
    v^T \mathbf{A} = \begin{pmatrix}\mathcal{C}_s^{[1]} &-\mathcal{C}_u^{[1]}\end{pmatrix} \begin{pmatrix}A_{ST} \\A_{TU}\end{pmatrix} =0\implies \mathcal{C}_s^{[1]}(A_{ST}) = \mathcal{C}_u^{[1]}(A_{TU}). 
\end{equation}
One can check that the color ordered gluon correlators in eq \eqref{eq:coloredYM} satisfy the above differential relation.

\section{Differential Representation of scalar with gluon exchange}\label{section5} In this section, we give a differential representation of four-point function of scalars mediated by a gluon. Our result closely follows the result of \cite{Herderschee:2022ntr} in the position space, see \ref{sec:AdS-review} for a brief review of this construction. 
Let us consider the seed as the scalar contact term,
\begin{equation}
    \Phi_4^{(0)} = \frac{1}{\Sigma}, \quad \Sigma = S+T+U
    \label{eq:seed}
\end{equation}
and define the following differential operators,
\begin{equation}
    \hat{n}_s = \mathcal{C}_t^{[0]}-\mathcal{C}_u^{[0]},\quad \hat{n}_t = \mathcal{C}_u^{[0]}-\mathcal{C}_s^{[0]}, \quad \hat{n}_u = \mathcal{C}_s^{[0]}-\mathcal{C}_t^{[0]}
\end{equation}
Here $\mathcal{C}_i$ refers to a Casimir for the $i^{th}$ channel,
\begin{equation}
    \mathcal{C}_{s} = \mathcal{C}_{12}, \quad \mathcal{C}_{t} = \mathcal{C}_{14}, \quad \mathcal{C}_{u} = \mathcal{C}_{13},
\end{equation}
where
\begin{align}\label{eq:Casimir12}
    \mathcal{C}_{12}[f] =& -5 \partial_{12}(c_{12}f) - c_{14}c_{23}\partial_{24}\partial_{13}f - c_{13}c_{24}\partial_{14}\partial_{23}f -\partial_{13}\partial_{23}(c_{13}c_{23}f) - \partial_{14}\partial_{24}(c_{14}c_{24}f) \notag\\
    &+ \sum_{ij\in\{12,13,14,23,24\}}\partial_{12}\partial_{ij}(c_{12}c_{ij}f)+(c_{13}c_{24} - c_{14}c_{23})\partial_{12}\partial_{34}f - 2(l_1^2 + l_2^2 - 2)f
\end{align}
Similarly, the expression for $\mathcal{C}_{13}$ and $\mathcal{C}_{14}$ can be obtained by doing a $2\leftrightarrow 3$ and $2\leftrightarrow 4$ exchange, respectively. Since we will be interested in identical external spins, we take $l_1 = l_2 = l$, and thus the index $\mathcal{C}_{i}^{[l]}$ refers to the external leg spin.

\noindent Here, note that the differential operators $\hat{n}_i$ satisfy the Jacobi identity trivially
\begin{equation}
    \hat{n}_s + \hat{n}_t + \hat{n}_u = 0
\end{equation}
From the action of Casimir $\mathcal{C}_i^{[0]}$ on the scalar seed,
\begin{equation}
    \mathcal{C}_{i}^{[0]}\Phi_4^{(0)} = 2(1+x_i)\Phi_4^{(0)}, \quad \textrm{where}~~ x_s = \frac{S}{\Sigma}, x_t = \frac{T}{\Sigma}, x_u = \frac{U}{\Sigma}
    \label{eq:cas-action}
\end{equation}
the four-point function for external scalars with a gluon exchange (S-channel) is given by,
\begin{equation}
    A^{s}_\gamma = \frac{\hat{n}_s}{S}\Phi_4^{(0)} \simeq \frac{1}{\Sigma}\frac{T-U}{S}\Phi_4^{(0)}
\end{equation}
and similarly for other channels, where $\simeq$ means up-to an overall numerical factor.\footnote{Similarly, we can get the four-point scalar  correlator with  the graviton exchange through the scalar seed $\Phi_4^{(0)}$ \eqref{eq:seed} (s-channel) as
\begin{equation}
    A^s_{g} = -(\mathcal{C}_s^{[0]}+6)^{-1}\hat{\mathcal{L}}_s\Phi_4^{(0)}
\end{equation}
where $\hat{\mathcal{L}}_s$ is a quadratic operator,
\begin{equation}
    \hat{\mathcal{L}}_s = \frac{3}{2}(\hat{n}_s)^2 - \frac{1}{2}(\mathcal{C}_s^{[0]})^2 + 10 \mathcal{C}_s^{[0]} - 24
\end{equation}
Using eqn. \eqref{eq:cas-action}, it's action on the scalar seed is given by
\begin{equation}
    \hat{\mathcal{L}}_s\Phi_4^{(0)} = 16 x_s^2 P_2\left( \frac{x_u - x_t}{x_s} \right)\Phi_4^{(0)}
\end{equation}
where $x_s,x_t$ and $x_u$ are defined in eqn. \eqref{eq:cas-action}. Thus, giving the s-channel four-point function to be
\begin{equation}
    A_g^s \simeq \frac{1}{(S+T+U)^2}\left( \frac{(U-T)^2}{S} - \frac{S}{3} \right)
\end{equation}}

\section{Discussion and Outlook}\label{section6}
Our analysis shows that the Grassmannian provides a simple and unified framework for studying both algebraic structures, such as color-kinematics duality, and differential relations, such as BCJ relations. This is a novel feature of the cosmological Grassmannian framework: unlike earlier approaches formulated purely in momentum space or embedding space, it realizes all these structures. We also obtained an exact kinematic Jacobi identity, i.e. $n_s+n_t+n_u =0$ in Grassmannian, which has not been achieved clearly from the AdS side \cite{Armstrong:2020woi,Albayrak:2020fyp}.\footnote{This should be contrasted with momentum-space Witten-diagram approaches to Yang-Mills theory in curved space earlier, where the natural kinematic numerators do not generically satisfy the Jacobi identity away from the flat-space limit, although a color-kinematics-dual representation can be obtained through an appropriate generalized gauge transformation} 
This opens up some natural directions that would be interesting to pursue- 
 
 \paragraph{4D-Grassmannian}
 First, it would be interesting to investigate color-kinematics duality and differential BCJ relations in the recently developed four-dimensional Grassmannian formulation of conformal correlators \cite{Bala:2026trw, S:2026qwn}.
\paragraph{Higher Points and Loops} A second important direction is the extension to higher-point Grassmannian correlators. In flat space, color-kinematics duality and the double copy are naturally formulated at the level of loop integrands. It would also be interesting to extend the present construction to loop level. A construction of loop-level Grassmannian representation would be very important direction.


\section*{\Large{Acknowledgment}}
SJ would like to thank the organizers and participants of Holostrings III, held at IIT Kanpur, where some aspects of the results in this paper were discussed. AB acknowledges a UGC-JRF fellowship. We would especially like to acknowledge our debt to the people of India for their steady support of research in the basic sciences. 

\bibliographystyle{JHEP}

\bibliography{biblio}

\end{document}